# Challenges in Community Resilience Planning and Opportunities with Simulation Modeling

**Abbas Ganji**
University of Washington
ganjia@uw.edu

**Negin Alimohammadi**
University of Washington
ngn.alimohamadi@gmail.com

**Scott Miles**
University of Washington
milessb@uw.edu

**ABSTRACT**

The importance of community resilience has become increasingly recognized in emergency management and post-disaster community well-being. To this end, three seismic resilience planning initiatives have been conducted in the U.S. in the last decade to envision the current state of community resilience. Experts who participated in these initiatives confronted challenges that must be addressed for future planning initiatives.

We interviewed eighteen participants to learn about the community resilience planning process, its characteristics, and challenges. Conducting qualitative content analysis, we identify six main challenges to community resilience planning: complex network systems; interdependencies among built environment systems; inter-organizational collaboration; connections between the built environment and social systems; communications between built environment and social institutions' experts; and communication among decision-makers, social stakeholders, and community members. To overcome the identified challenges, we discuss the capability of human-centered simulation modeling as a combination of simulation modeling and human-centered design to facilitate community resilience planning.

**INTRODUCTION**

Community disaster resilience is defined as "the ability of a community to prepare for anticipated hazard, adapt to changing conditions and withstand and recover rapidly from disruptions" (NIST 2016: V.1:1). Resilience of community is built upon a variety of community aspects that have been extensively investigated in research studies (Aldrich and Meyer 2015; Comes and de Walle 2014; Chang and Rose 2012). As disasters may cause extreme damage and long-lasting disruptions in community functioning, it is vital for community stakeholders to envision potential damages and expected recovery process beforehand. For this purpose, several community resilience planning initiatives have taken place in the U.S. in the last decade to identify community hazards, anticipate the recovery processes, establish community resilience goals and provide recommendations for decision-makers to achieve better community resilience (OSSPAC, 2013; WASSC, 2012; Poland, 2009).

Community resilience is complicated and multi-dimensional on its own. Community resilience planning, in practice, is a highly collaborative process that involves numerous community stakeholders associated with human-oriented challenges. Participants in community resilience planning include emergency managers, experts and managers of building and infrastructure systems, social stakeholders, and so on. Community resilience planning is technical and domain-oriented and requires a comprehensive understanding of the recovery process of damaged entities in the built environment and social systems. In addition to being a complex puzzle, community resilience planning is a highly user-centered process, requiring collaboration among built environment experts with different areas of expertise, communication among experts of the built environment and social institutions, and information sharing among the planning participants, community stakeholders, governmental and elected officials, and community members. It is essential to recognize and understand the complexity of the planning process and its characteristics. As well, considering that several community resilience planning initiatives taken place in the U.S., it is beneficial to learn from these experiences to facilitate future initiatives.

In this study we investigate the process of community resilience planning and identify challenges that participants experienced in previous planning initiatives as our first research question. The next step is to determine how community resilience planning can be facilitated. Simulation modeling is widely used to model connectedness and interdependencies in network systems such as infrastructure systems (Ramachandran et al. 2015). Several research studies have successfully used simulation models to investigate community disaster recovery processes (Miles at al. 2018). However, no analytical computer-based tools like simulation models have been used in previous planning initiatives. As our second research question, we investigate if simulation models can facilitate planning initiatives, and if so, how they can address the identified challenges. Analyzing interviews conducted with experts who participated in the initiatives, we conclude that combining human-centered design with simulation modeling, referred as human-centered simulation modeling, has the potential to address observed challenges.

**BACKGROUND**

Community resilience is a community attribute that can be improved and adapted over time; it refers to the ability of a community to mitigate and resist against hazards and its ability to recover quickly (Bruneau et al. 2003). Several frameworks have been developed to describe the foundations of community resilience in different domains (Miles 2015; Kuling et al. 2013; Norris et al. 2008). As disasters impact various aspects of a community, community disaster resilience is also considered in diverse dimensions such as social institutions (Aldrich and Meyer 2015; Alipour et al. 2015; Semaan and Hemsley 2015), physical infrastructure (Cimellaro et al. 2011; Davis et al. 2018), economics and business (Chang and Rose 2012), and healthcare (Chandra et al. 2011; Comes and de Walle 2014). Berkes and Rose studied the characteristics of community resilience and integrated two dimensions of social–ecological systems, the psychology of development and mental health, in a framework (Berkes and Rose 2013). Chang et al. developed an approach to characterize communities' infrastructure resilience and identified key challenges including incomplete incentives and partial information (Chang et al. 2014). Berk et al. analyzed coastal state hazard mitigation plans and compared the quality of these plans (Berk et al. 2012). Labaka et al. (2014) presented a framework to identify resilience policies across technical, organizational, economic, and social dimensions (Labaka et al. 2014). Rubim and Borges (2017) conceptualized and characterized resilience in the context of complex systems (Rubim and Borges 2017). Turoff et al. developed a model for interaction among critical infrastructure systems (Turoff et al. 2016). Additionally, many research efforts have attempted to quantify community resilience and identify quantitative indicators and indices to evaluate different dimensions of community resilience (Cutter et al. 2016 and 2010; Bruneau et al. 2003).

Community resilience planning requires community decision-makers, built environment experts, and social stakeholders to collaborate to identify social goals and their dependencies (NIST 2016: V.1:1). This broad collaboration and communication among experts in different systems and non-experts makes community resilience planning more challenging due to involving human-factors in the planning process. Characteristics and challenges in emergency management have been extensively studied (Scholl and Carnes 2017; van Laere et al. 2017; Turoff et al. 2016; Gonzalez et al. 2012; Maitland et al. 2009; Dilmaghani et al. 2006). However, the existing literature does not specifically identify challenges in community resilience planning. In next section, we introduce three community seismic resilience planning taken place in the US in last decade. The initiatives are a valuable source for finding existing barriers in community resilience planning.

**COMMUNITY SEISMIC RESILIENCE PLANNING IN THE U.S.**

Bruneau et al. (2003) published one of the early research studies to define the concept of community seismic resilience (Bruneau et al. 2003). They defined community seismic resilience as "the ability of social units (e.g., organizations, communities) to mitigate hazards, contain the effects of disasters when they occur, and carry out recovery activities in ways that minimize social disruption and mitigate the effects of future earthquakes" (Bruneau et al. 2003). They also proposed a framework to quantify community resilience to be evaluable and measurable. This research and other studies led community experts to apply and quantify community resilience to real communities. Recognizing the importance of community resilience in the research studies, seismic committees and urban planning associations formed and organized planning initiatives to assess resilience of their own communities.

### SPUR Resilient City (2006-2009)

The first attempt at community seismic resilience planning was organized by the San Francisco Bay Area Planning and Urban Research Association (SPUR) (Poland, 2009). Experts from architecture, engineering, urban planning, and public policy planning firms were invited to work together to envision what would happen to the city after a high-magnitude earthquake. In this initiative, participants defined the concept of disaster resilience, selected a single city-wise expected earthquake scenario (10% occurrence in 50 years) as the community hazard level, established desired timeframes of recovery for buildings and infrastructure systems, estimated the anticipated resilience state for identified clusters of buildings and infrastructure systems, and provided recommendations for elected officials and community stakeholders. Comparing anticipated and desired resilience states identifies the gap between where the community is and where it should be. This concept has become more popular in similar studies. It should be noted that the current (anticipated) recovery timeframes were collaboratively estimated based on experts' judgment. This work inspired seismic committees in other states, and the procedure was followed by other community seismic resilience planning initiatives with some modifications.

### Resilient Washington State (2010-2012)

In 2010, the Resilient Washington State (RWS) subcommittee of the Seismic Safety Committee under the Washington State Emergency Management Council launched a community seismic planning initiative for the state of Washington inspired by SPUR (WASSC, 2012). Since RWS held a statewide initiative, they referred to National Seismic Hazard Maps created by the U.S. Geological Survey (USGS) in 2008 to identify community hazard. RWS formed four groups consisting of critical services, utilities, transportation, and housing & economic development sectors, identified their components, and located experts from these sectors. They followed the SPUR procedure, identified community hazard, and presented desired and anticipated recovery timeframes for community entities. They also provided several recommendations for community policy makers and published their report in 2012.

### Oregon Resilience Plan (2011-2013)

The Oregon Resilience Plan (ORP) was conducted as a community seismic resilience planning initiative in 2011-2013 by the Oregon Seismic Safety Policy Advisory Commission (OSSPAC) (OSSPAC, 2013). They considered a M9.0 Cascadia earthquake and tsunami as the community hazard for the state of Oregon. ORP consisted of eight task groups: Cascadia Earthquake Scenario, Business and Workforce Continuity, Coastal Communities, Critical and Essential Buildings, Transportation, Energy, Information and Communications, and Water and Wastewater. The planning process was similar to that of SPUR and RWS, but more detailed and specific to the state of Oregon. Compared to the previous initiatives, ORP perspicuously appointed long-term community goals and social needs. They established desired recovery timeframes of community entities such that business continuity in the state would not be disrupted for more than two weeks to one month. The task groups considered this goal as a criterion to identify desired recovery timeframes. As a result, the existing gaps between the anticipated state and desired state were well understood by community decision-makers and the initiative's audiences.

### Community Resilience Planning Guide for Buildings and Infrastructure Systems by NIST (2016)

The National Institute of Standards and Technology (NIST) published a two-volume report, *Community Resilience Planning Guide for Buildings and Infrastructure Systems*, in 2016 examining these initiatives (NIST, 2016). This report brought community resilience planning to the forefront, with significant updates and details. Although NIST's report offered similar reasoning and conclusions as the initiatives themselves, it also provided valuable details on **how to proceed** with systematic community resilience planning. The NIST report identified community representatives who should be involved in planning. It also divided communities into the built environment dimension, including buildings and infrastructure systems, and the social dimension consisting of social institutions such as businesses, industries, and financial systems, and recommended identifying the characteristics of each dimension. More importantly, NIST suggested linking social functions with the built environment, a concept that was not explored in the existing initiatives. In this way, the relationships among numerous systems, in either the social or built environment, are well defined and the community is envisioned as it really is—an integrated whole rather than separate groups that function individually and independently. The NIST report highlighted dependencies and cascading effects, and presented a dependency matrix among infrastructure systems, which had not been explored by the three initiatives although they had noted its significant impact.

Community resilience planning has evolved based on the lessons learned from the planning initiatives, actual events, and research studies. Similarly, the process of resilience planning has been enriched by taking into account more influential concepts, parameters and dimensions, and defining relationships among them. In summary, NIST suggested following the steps presented in Figure 1.

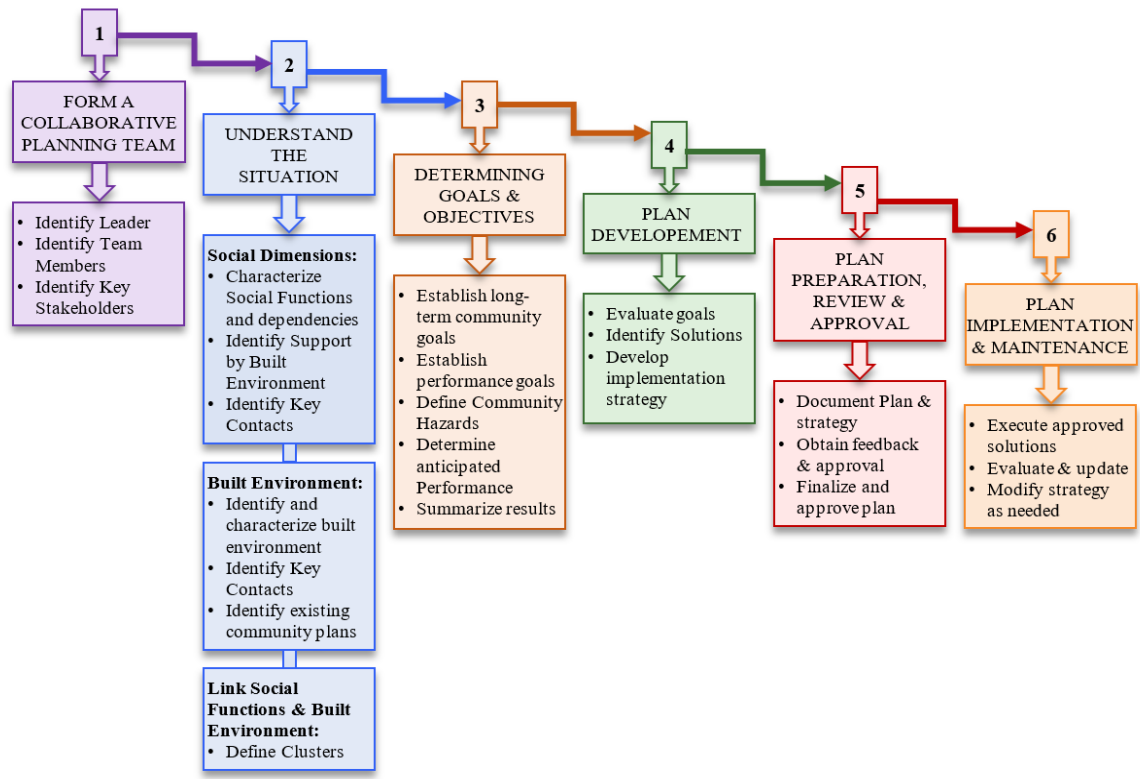

**Figure 1. Six-step planning process for community resilience (NIST, 2016)**

Although community resilience planning has grown quickly and NIST's report presented it as a standard and unified procedure, undertaking planning in practice comes with wide variety of surprises and challenges. In this study, we look for challenges that participants in the community resilience planning initiatives experienced. We also discuss possible ways to overcome these challenges.

**METHODOLOGY**

As mentioned earlier, we examine two research questions in this study. First, we aim to identify challenges appeared in the community resilience planning initiatives, and second, we want to investigate how simulation modeling can help this process from the participants' perspective. For this purpose, we reviewed the procedures and conclusions of three initiatives and conducted semi-structured interviews with the initiatives' participants to learn about the process and observed challenges. Details on the interviewees, data collection, data coding and analysis approach are provided below.

**Participants and Data Collection**

The first author of this study contacted several critical infrastructure agents, emergency managers, and researchers who participated in one of the initiatives via email, described the purpose of the interview, and requested a 90-minute in-person meeting to learn about their experience in the initiatives. He invited interviewees from different initiatives to recognize similarities and differences. He also interviewed a manager of critical infrastructure system who did not participate in any of the initiatives but has conducted several relevant research studies and community resilience discussion meetings with experts in infrastructure systems.

This interview was to include the perspective of an independent expert who did not participate in any of the initiatives. Overall, the first author conducted 90-minute semi-structured interviews with eighteen emergency managers, infrastructure and building experts, and researchers. Seventeen interviews were conducted in-person, and one interview was over the phone. Geographically, the interviewees were from California, Oregon, and Washington (five, seven, and six interviewees from each state, respectively). All interviews were recorded and transcribed with interviewee's permissions for further analysis except one interview where the interviewee did not grant permission to record the conversation. We relied on hand-written notes for that interview and included these notes in our qualitative content analysis. Interviews were voluntary and unpaid, and the interviewees were kind and welcoming. Interviews were conducted from May to September 2018.

**Data Analysis and Qualitative Coding**

Our data analysis was driven by open-coding thematic analysis. The first author applied line-by-line coding to five random interviews before establishing a codebook. Initial codes emerged from finding relationships based on the line-by-line coding. The third author, who is an experienced social scientist, university researcher, and expert in community resilience, inspected the initial codes. By establishing the initial codebook, the first and second authors coded the entire dataset. For collaborative qualitative coding, we used Code Wizard as a collaborative coding tool (Ganji et al. 2018). Code Wizard consists of programmed Microsoft Excel spreadsheets. Code Wizard is free, appropriate for small to midsize teams and academic research projects, and does not require much training, which was a significant concern for the authors. The first and second authors coded data independently and separately. Using Code Wizard, after each round of individual coding, we aggregated coded data, evaluated the Inter-Coder Reliability (ICR) coefficient, noted problematic codes, discussed the reasons for disagreement, and revised the codebook. We performed three rounds of coding to meet an acceptable ICR threshold (0.8). We coded data for two themes: (1) challenges and (2) opportunities for simulation modeling.

**CHALLENGES IN COMMUNITY RESILIENCE PLANNING**

Based on our qualitative content analysis, we present the main challenges in community resilience planning identified by the interviewees. We used the term **main** challenges because there were additional challenges mentioned in the interviews that we intentionally ignored. For example, a few issues occurred only in one of the initiatives and were not generalizable for other initiatives, and some other barriers taken place due to the logistics of holding meetings and time limitations.

**Complex Network Systems**

According to NIST, built environments that support community members and social community functions consist of infrastructure systems and buildings. Recovery of an infrastructure system to be able to provide services to consumers, regardless of its interdependencies with other systems, depends on two factors: (1) recovery of discrete damaged entities, and (2) the connections and dependencies among networked entities. Both factors should be considered in planning since ignoring either one results in a distorted and incomplete picture of the recovery process of infrastructure systems.

In the planning initiatives, participants formed task groups for infrastructure systems, and participated experts joined the groups based on their areas of expertise. Experts identified the main entities of each sector in the task group. Through technical discussions and aggregating experts' judgment, the task groups created anticipated and desired recovery timeframes for identified entities (see Figure 2). Experts typically estimated damages and recovery timeframes based on community hazard intensity, liquefaction and landslide potentials, and type, material, and age of entities.

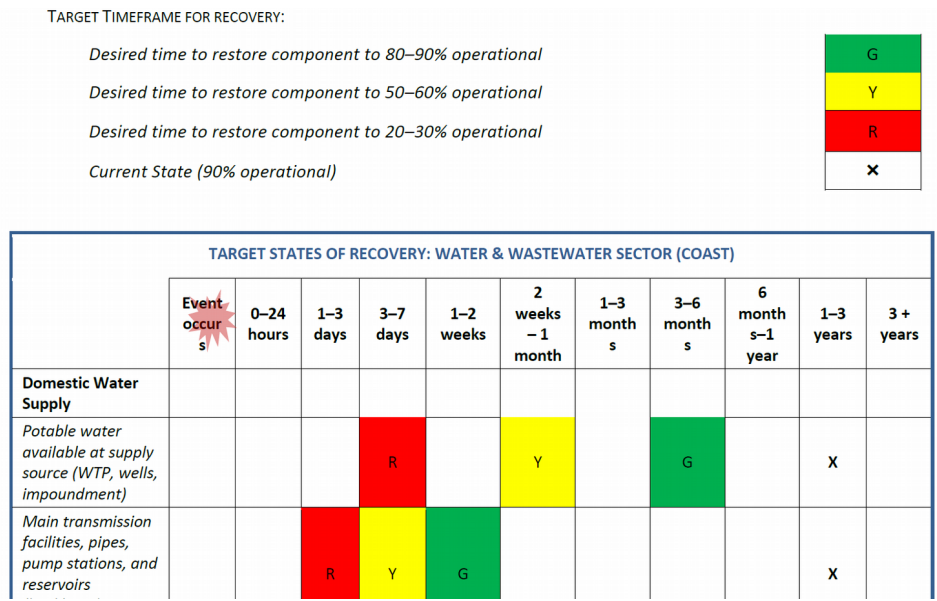

**Figure 2. Desired and Anticipated Recovery Timeframe in Water & Wastewater Sector (OSSPAC, 2013)**

Experienced experts provided insightful information to estimate required time for recovery of single entities. This information can be used as a basis in community resilience planning. However, overlooking the network and interconnected structure of infrastructure systems underestimates recovery timeframes from the perspective of community members who need these services regardless of the recovery of single entities. System resilience depends on the level of redundancy that exists in the system. This redundancy is gained from the network system and its structure, not just entities.

> [Interviewee #7:] "…in water systems, there's so much dependency that exists, you've got to restore this before you restore that, before you restore that, and similarly in capacity… if you're going to restore the distribution system, you have to have sufficient supply to be able to meet that, but there's no need to build the supply capacity any faster than you can build the delivery system to be able to meet those needs."

Due to lack of appropriate tools, and limited time and budget, the network and dependencies in infrastructure systems were not considered in the planning initiatives. In order to make anticipated recovery timeframe estimates more realistic, the network system should be taken into account in future studies.

**Interdependent Network Systems and the Cascade Effect**

A damage in an infrastructure system may result in disruption of serviceability of many other systems. This phenomenon, referred to **cascade effect,** is also observed in the recovery process of interdependent systems. For example, the water system relies on electricity for running pump stations and back-up generators. Therefore, serviceability of the water system depends on not only the functionality of the water system itself, but on the power system as well. Similarly, recovery of both water and power systems is contingent on access to damaged entities, which makes recovery process of water and power systems dependent on recovery of transportation system. The significant impact of interdependencies among infrastructure systems and buildings was noted in the planning initiative reports and discussed in more detail in the NIST report. However, interdependencies have not been taken into account for estimating anticipated recovery timeframes of built environment in the planning initiatives.

> [interviewee #10:] "there are significant interdependencies. It's hard to really comprehend to what level they are important. To me, power is the big one. What can happen until power is restored? Not a whole lot...you probably can have some restorative construction happen without power, but even that's a challenge. Everything relies upon power, and how resilient is the power. What are those interdependencies, does that back everything else up in recovery?"

**Inter-Organizational Collaboration**

In addition to the domain-oriented characteristics presented above, community resilience planning is challenging due to its inter-organizational nature. For example, infrastructure systems and buildings are managed by different agencies, companies, and organizations with dissimilar procedures, priorities, and goals. Even a single infrastructure sector might be managed by multiple companies and decision-makers. Planning to make a community resilient entails all decision-makers from different organizations coming to agreement on how to proceed and what goals to pursue. Time needed for a community to recover after a disaster depends on the efficiency of experts' collaboration. For example, achieving desired recovery timeframes requires experts and managers in companies that are part of the power system (if there are multiple companies providing electric service for the community) to know how to plan for resilience to achieve these goals and how different policies might affect resilience of the entire community. Therefore, inefficient and ineffective inter-organizational collaboration is a big challenge in community resilience planning.

> [Interviewee #2:] "In the power system, for example, there are multiple power system providers and they each have their own proprietary information, and probably many of them have done studies. [A company in infrastructure system] has studied some of their things, [another company in infrastructure system] has probably studied some of theirs, maybe they haven't studied everything, but a real challenge is getting access to that information, for starters."

Similarly, interviewee #4 explains other challenges that emerge when decision-makers and stakeholders are from the public and private sectors and comply with different regulations:

> [Interviewee #4:] "[infrastructure] is public and private, that's the important thing. Almost none of it is purely public, and so [...] there is proprietary information that those private sector companies hold close and have a right to hold close. And so, there are regulatory agencies, there's the utility and transportation commission, and they will honor their right to proprietary information. They try to work out planning with them, but some of it, for instance, even though you might say you want the hospital to come back first, [...] depending on the way the power outage happened, it may be that you have to fix something somewhere remote to there, because you also have to figure out how to bring the system back, and just bringing back one small branch of the system, even if it does feed the most vulnerable sector, may not do you the most good if bringing back a major substation gets you the most bang for your buck in the first twelve hours. And how they make that decision, again, is private."

**Connections between Social Institutions and Built Environment**

Citizens and community members found social institutions to address their needs. NIST categorized social institutions in eight groups: (1) family/kinship; (2) the economy; (3) government; (4) health; (5) education; (6) community service organizations; (7) religious, cultural, and other organizations that support belief systems; and (8) the media. While the goal of community resilience is closely tied with the recovery of social institutions, these institutions themselves rely on functionality of the built environment (including infrastructure systems and buildings). Consequently, desired recovery timeframes for the built environment are established based on social needs. Social and built environments are connected in community resilience planning and it entails identifying "links between social functions to the supporting built environment" (NIST 2016: V.1:8). The importance of identifying these links has been gradually recognized in the community resilience planning initiatives.

The NIST report devotes an entire chapter to identifying these links. However, the three initiatives, while recognizing the importance of identifying these links, did not really consider these links to estimate anticipated or desired recovery timeframes. NIST and the planning initiatives present desired and anticipated recovery timeframes for the built environment by percentages of functionality of systems entities. For instance, the desired recovery time for functionality of 80-90% of distribution pipes is determined 1-3 days in Washington state (WASSC, 2012). The desired time to restore 50-60% of electric transmission lines in the "non-tsunami coast zone" in Oregon was determined 1-3 weeks (OSSPAC, 2013). However, it is not clear how these timelines have been assigned to support social institutions and community members where connections between the built environment and the social dimension are lacking. As another example, the ORP defines the desired recovery timeframes of infrastructure systems such that businesses, as a social entity, would not experience disruption in having infrastructure services longer than two weeks. Imagine a hypothetical scenario in which 60% of electric transmission lines have been recovered in two weeks (determined as desired recovery timeframe in ORP), but business institutions are relied on the 40% unrecovered systems. In other words, although the desired recovery timelines for the built environment may be met, it is not certain whether or not businesses would receive services from the built environment in two weeks.

> [Interviewee #6:] "I think the other biggest shortcoming was there wasn't the opportunity to go through and say, "here is an integrated set of recommendations drawn from each of the areas that together get us the furthest down the line to resilience and … address the question of what's the appropriate sequencing and balancing of these things.""

It should be noted that connection and dependence between the social dimension and built environment is mutual. While we described how recovery of the social dimension depends on the built environment, recovery of the built environment also relies on social dimensions. For example, recovery of infrastructure systems or buildings is executed by experts, crews, engineers, and infrastructure managers. As community members, they need social institutions such as healthcare and finance/economic institutions to be able to first survive and then participate in the recovery process.

> *[Interviewee #7:] "… their experience [New Orleans Water and Sewer Board] was so revealing to me. When [Hurricane Katrina] happened, they had all of this flooding, they had all of these facilities out. More than half of their workforce did not come back to work, and so how can you restore the system and establish service to the community if you don't have the workforce necessary to be able to do anything more than basically keep the lights on? […] I can't just take just anybody and put them on a backhoe and tell them how to repair a pipe. I need trained people who are able to respond to this event and aid in the restoration and recovery of our water system. […] The other thing I want to leave you with is it goes way beyond just infrastructure. I touch on obviously the people part of it, and that's key, but the business operations side of what we do is equally important, and again, water is a wonderful example as a utility, but it's true with most every other lifeline with the exception of maybe transportation. Somebody has to pay the bill. And so, if we're out of business, we're not sending out bills, we're not collecting revenue, we're not paying out bills, and pretty soon… the good news is I can restore the water system, but I can't pay for the fuel because nobody will take my credit card. I can't pay the electric bill to run the pumps that I need, and so I don't have the ability to run a business, to be able to sustain this organization, to cut paychecks to the people that I need so desperately to be able to repair the pipes."*

**Communication and Information Sharing between Experts in Built Environment and Social Institutions**

Built environment experts such as managers, agents and decision-makers in infrastructure systems and social stakeholders involved in the planning must work together closely. As mentioned earlier, NIST states that decision-making for recovery planning is based on social needs and community goals. Since social dimension like social institutions depend on the built environment to be functioning, experts on the both sides must communicate to share required information. In addition to identifying physical connections and dependencies between these dimensions, collaboration and communication between stakeholders in these two groups are challenging for community resilience planning.

> *[Interviewee #13:] "So, as we move into thinking about resilience, which is about more than the structure, it's about the occupancy, it's about more than one building, it's about the organization like a community. We have to be open to working with other groups, and this is a challenge, because those other groups never heard from us, in fact they're much bigger than us. They don't want to really hear engineers, and they have almost no patience."*

One important aspect of this collaboration and communication is prioritization, which is challenging step in community resilience planning. Prioritization is inevitable due to time and resource limitations, and potential conflicts in community recovery. Communities have similarities but are completely different in terms of their built environment structures and social needs and cultural differences. Therefore, as NIST concludes, there is no global priority list or gold standard for decision-making in community recovery planning, and priorities should be determined in a community-to-community basis. It underlines the significance of needs to strengthen collaboration, communication and information sharing among decision-makers from any dimension.

> *[Interviewee #7:] "... there are competing needs for limited resources. How as a community do we prioritize those limited resources to be able to meet the needs of the recovery, to be able to get recovery balanced and do it in a way that limits loss of life and expedites the overall recovery process? This is so challenging."*

**Information Sharing and Communication between Experts and Community Members**

This category of challenges in community resilience planning came as a surprise in our interviews. The final audience of community resilience planning initiatives is members of the community and their elected officials. Executing the decisions made in community resilience planning is a very costly and lengthy process that requires the support of the community. Experts in community planning share information with community decision-makers, elected officials, community policy makers, and finally community members. They aim to inform community members and decision-makers about the hazards that threaten their community and warn them about potential consequences of such disasters. They also envision the community's preparedness to resist and bounce back after the expected disaster and point out the gaps between the current and desired states of preparedness. Making a community resilient requires a budget and resources which are usually obtained from taxpayers. Resilience planning cannot successfully proceed if community members do not properly understand its importance and support long-term investments in it.

> *[Interviewee #10:] "So, [this is] all state taxpayers' money going in to support this grant program to get this money out. Some of the successes and challenges associated with getting the grant program up and running, included people who didn't understand why it was important to have safe schools... schools' job [is to] educate, not to do seismic mitigation of schools. So, there was a really big learning curve for [...] really a lot of people."*

Considering the notes above, communication with the wider community is challenging because it has its own complexities and requires specific types of skills, especially when people have incorrect perceptions about the likely post-disaster condition.

> *[Interviewee #6:] "I think it was not well understood on the part of the public... [T]hat was one of the most important things that the resilience plan did was really put some parameters on it that the public could access, and I think that a lot of the public perception in the past had either been "oh, it's not going to be that big of a deal, I don't need to worry about it" or "it's going to be the end of the world, so why should I worry about it, there's nothing I can do, everything is going to be completely destroyed." Some of the commentary that came out from Tohoku, and some of perceptions that people got from the New Yorker article kind of fueled that notion of "oh my god, it's going to be this total disaster and nobody's going to survive, there's no point in preparing." And the resilience plan, I thought, the clearest message of it was "pretty much everybody is going to survive, and you're going to get up the next morning and not going to be able to flush the toilet for three months. That's going to be your problem." That's what we have to address, is that massive disruption to everyday life that is going to make it very difficult to live here, and certainly going to make it very difficult to continue to be employed here. And so … although at the time I felt like we're just doing this, engineers are just kind of guessing in terms of the damage and they tend to guess conservative when put in those kinds of situations, I think it still ended up with the right overall message which is "this is going to be really difficult, but there are things you can do to solve, to fix it, and move ahead with it."*

**FACILITATING COMMUNITY RESILIENCE PLANNING BY SIMULATION MODELS**

Reviewing the reports published after the planning initiatives and based on the interviews, we noticed that other than Hazus, which was used limitedly for damage and loss estimations, no analytical computer-based tools have been used in any planning initiatives. For example, estimating anticipated recovery timeframes of damaged community entities are computationally challenging tasks in the planning initiatives. However, use of analytical computer-based tools such as simulation models was lacking in the planning process.

Simulation models are nowadays used in numerous fields such as engineering, urban planning, and supply chain management. The capability of simulation modeling to improve different phases of emergency management is widely recognized in research communities like Information Systems for Crisis Response and Management (ISCRAM). Similarly, in practice, Federal Emergency Management Agency (FEMA) emphasizes the importance of tools in emergency management, noting that "innovative models and tools" are one of three strategic needs to accomplish recovery planning (FEMA, 2012). Simulation modeling is also used in community resilience. The interviewees were asked about how simulation modeling could help them in community resilience planning. It should be noted that most of the interviewees had experience or knowledge about simulation modeling and its capability in community resilience planning. For other interviewees who had no experience, background or information in using simulation models in this field, we shortly explained how simulation modeling works and provided some examples from the literature. We analyzed their answers and categorized their responses based on the codebook that we had established to identify the challenges since (1) we could maintain the consistency of presented concepts for our audiences, and (2) we could benefit from the details provided by interviewees to understand what tasks and functionalities are needed to handle by simulation modeling in this process. The interviewees' responses reveal what limitations and challenges can be addressed by simulation modeling and where simulation models can facilitate them. In the following, we present the benefit of using simulation modeling to address each category of identified challenge.

**Complex Network Systems**

As discussed previously, infrastructure systems are complex networks. In such complex systems, it is extremely difficult to predict how recovery proceeds and when these systems would be able to provide services to others, including community members, social institutions, and other dependent infrastructure systems. Simulation modeling has the capability to consider this complexity, and many research studies have used simulation modeling to simulate the post-disaster recovery process of infrastructure systems and housing as mentioned in background section. The interviewees also mentioned that simulation modeling would be helpful in community resilience planning.

> *[Interviewee #5:] "I believe if there were modeling tools available to estimate downtime, the connectedness, and the dependencies of the different lifeline sectors, … and how doing certain fixes would improve the rapidity of getting things up and running quickly, I think it would help with building resilience and mitigation projects."*

Simulation modeling also makes planning much easier if experts would like to update data. If a simulation model is already set, new updates in system entities taken place over time can be simply applied to the model to adjust outputs accordingly.

> *[Interviewee #8] "And a tool would be nice to have, something where [... it] allows you to improve the data. So, you can put in with what you know, understanding your margin is bigger, but as you get better data, that comes around [...] that can be updated and give you better results."*

> *[Interviewee #12] "we need to figure out the things that we can't work around after the disaster ... that's what the days, weeks and months are about, to identify how much time they can have to get something going again but do that kind of simulation so that you can understand what you can work around."*

**Interdependencies Among Infrastructure Systems**

The advantage of using simulation models to capture the complexity of system interdependencies in community resilience planning was mentioned by the interviewees more than other simulation modeling capabilities. Simulation modeling has been widely used to simulate interdependencies among critical infrastructure systems in urban planning.

> *[Interviewee #7:] "simulation modeling would really be critical to illustrate this interdependency effect … and again, I had not noodled through it, everything, when I give this talk at many different seminars, I talk about interdependencies and how everything is connected to everything."*

> *[Interviewee #14:] "If there was a way to facilitate the collaboration over the issue of interdependence, right, if there was a way to synthesize the data that reflects the way one system depends on another, and then use that to run scenarios."*

> *[Interviewee #2:] "I think, what you're talking about [possibility of using simulation modeling], in terms of being able to model those interdependencies, [is] very valuable, and then being able to communicate that, again, to the community, to our power providers, and having redundant systems, redundant load paths to bring power into the community, that's the ultimate goal."*

**Inter-Organizational Collaboration**

Planning for community resilience is highly collaborative. Experts need to collaborate with each other since organizations have their own plans, policies, and priorities, but their decisions impact other groups' planning. This collaboration is difficult to facilitate when several organizations are involved; this is another area where simulation modeling can be used, as it can facilitate information sharing among experts.

> *[Interviewee #11: ]"So the kinds of things that I want to see [if I had simulation models] are probably more rooted to other infrastructures, you know, the buildings, roadways, bridges [...] if I had to then bring resources in, it would also give me an idea of what I might be able to work with there. [...] But if I can't get from here to there, if there's something blocking, today I can't plan on that."*

**Connecting Social Institutions and Built Environment**

As discussed earlier, a significant challenge in community resilience planning is connecting built environment entities and social institutions because they mutually rely on and support each other. This is extensively described by NIST due to its critical impact on planning. However, these connections and links were not identified and considered in detail in the planning initiatives due to their extreme complexity. Simulation models can be incorporated to facilitate community resilience planning in this regard. The interviewees also mentioned this and provided some details of what they expect from simulation models to help them.

> *[Interviewee #8:]* "If you could have a model and say "what would be the outcome if we fixed up the road system first" or what would be impacted, any buildings, there would be a lot of URM [Underrepresented Minority]. You know, what would be our payback for looking, solving, retrofitting all of our URMs versus… To be able to ask some of these questions, when these things come up for planning purposes, this is where we really ought to be …"

**Communication between Built Environment and Social Institutions Experts**

Experts in built environment and social institutions need to communicate and share information to make sure that decision-making comprehensively considers both dimensions and their limitations. Simulation modeling can improve this communication by simulating recovery processes based on different scenarios and providing the consequences of these scenarios for experts to analyze. Since community resilience deals with many dimensions in a community, it is likely to overlook consequences that are critically important for some experts or decision-makers. Simulation modeling helps experts to quickly and inexpensively see how the recovery process changes based on their decisions.

> *[Interviewee #14:]* "just thinking about this idea of scenario-based thinking, if there's a way to help people sort of wargame the way these things play out. "Here's what it looks like now. We wargame. We run an exercise. We get some results. We make changes. We do it again. Hopefully we're even closer. By seeing what changes, we need to make, we're getting closer." So that kind of simulation might be intriguing."

As well, simulation modeling can help prioritize decision-making. It can provide timelines, expenses, and, more importantly, social impacts of decisions made. As a result, decision might be changed if inappropriate consequences threaten community values.

> *[Interviewee #8:]* "One of the questions that there's no really good tool for saying, for doing scenario planning. [...] you accepted that this is risk, so what do you do with all of that? So, how much does it cost you to do all of that stuff, over what amount of time? Where should I put my dollars? What should I do first? What should I do second?"

> *[Interviewee #2:]* "So, we're just coming up with a time, but what's not in there, I always call this the five percent problem, the standard for restoration is always ninety-five percent, but who is the five percent that doesn't get the power, and that we stop caring once we're at ninety-five percent? I want to know, you know, because I did the study in [a community in the U.S.] where I found that Hispanic populations literally were restored slower than white populations."

**Information Sharing and Communication between Experts and Community Members**

Effective communication and information sharing between experts and non-expert community members, or non-expert elected officials, is necessary in community resilience planning. Study participants and interviewees drew attention to the benefit of simulation modeling in this regard and mentioned how it can improve this process.

> *[Interviewee #7:]* "I think it [community resilience planning] depends on what the community's values are going to be, but it's a tremendous opportunity to have that conversation in advance [...] a tool like what you're talking about in terms of simulation allows conversation with the board or the public or the decision makers to happen in advance, so that when the event occurs you've got some guidance in terms of how to work"

> *[Interviewee #7:]* "More importantly, in many respects, as you go through this system planning process and the investments in the infrastructure over time to build the hardened backbone and harden the facilities, your system investments are consistent with those values and the ability to make the restoration in that kind of priority sequence."

**DISCUSSION AND CONCLUSION**

In this study, we identified the challenges associated with the previous community resilience planning initiatives taken place in California, Washington and Oregon. For this purpose, we interviewed experts, critical infrastructure managers, and emergency management agents who participated in the initiatives. Due to the capability of simulation modeling to address the challenges mentioned in the research studies (Ganji and Miles 2018; Miles et al. 2018), we asked the interviewees to share their opinions about how simulation modeling would support the planning initiatives and presented the findings in the previous section.

As mentioned earlier, simulation modeling is widely used in community resilience, emergency management and disaster recovery. Simulation modeling has been evolved through various approaches including (1) resource-constrained modeling, (2) machine learning, (3) dynamic economic impact modeling, (4) system dynamics simulation, (5) agent-based simulation, (6) discrete-event simulation, (7) stochastic simulation, and (8) network modeling (Miles et al. 2018). Several research studies used simulation modeling in disaster recovery of various elements of communities such as water and wastewater system (Tabucchi et al. 2010), power systems (Çağnan et al. 2006) and housing recovery (Longman and Miles 2019; Burton et al. 2017; Miles and Chang 2011). It shows the capability of simulation modeling to capture complexity of these network systems mentioned as the first challenge by the interviewees. The next challenge, the interdependencies among critical infrastructure systems, is also addressed in the literature of urban planning (Turoff et al. 2016; Banuls et al. 2013; Ouyang 2014). Integrating the built environment including buildings and critical infrastructure systems and social institutions, the forth challenge, is applied in simulation modeling of disaster recovery process (Ganji and Miles 2018).

While simulation modeling can address the first, second and forth challenges, other challenges are user-oriented and human-factors are involved. Human-centered design has the potential to address the user-oriented challenges. Human-centered design considers the concerns, values, and perceptions of all stakeholders in design of simulation models in a problem-solving process (Baxter and Sommerville 2011). Ganji and Miles (2018) presented a conceptual framework of characteristics of disaster resilience and recovery planning in two main categories including domain- and user-oriented characteristics (Ganji and Miles 2018). They argued how combining human-centered design and simulation modeling is capable to overcome not only domain-oriented challenges, but also user-oriented challenges, and proposed a conceptual framework for such models. We refer this combination as *human-centered simulation modeling*.

**ACKNOWLEDGMENTS**

The authors would like to thank Adele Miller for her effort in transcribing the interviews, data preparation and data cleaning. Funding support for this paper was provided by National Science Foundation Award No. 1541025.

**REFERENCES**

Aldrich, D. P., and Meyer, M. A. (2015). Social Capital and Community Resilience. *American Behavioral Scientist*, 59(2), 254–269.

Alipour, F., Khankeh, H., Fekrazad, H., Kamali, M., Rafiey, H., & Ahmadi, S. (2015). Social issues and post-disaster recovery: A qualitative study in an Iranian context. *International Social Work*, *58*(5), 689-703.

Burton, H. V., Deierlein, G., Lallemant, D., & Singh, Y. (2017). Measuring the Impact of Enhanced Building Performance on the Seismic Resilience of a Residential Community. Earthquake Spectra, 33(4), 1347-1367.

Banuls, V. A., Turoff, M., & Hiltz, S. R. (2013). Collaborative scenario modeling in emergency management through cross-impact. *Technological Forecasting and Social Change*, *80*(9), 1756-1774.

Baxter, G., & Sommerville, I. (2011). Socio-technical systems: From design methods to systems engineering. Interacting with computers, 23(1), 4-17.

Berkes, F., & Ross, H. (2013). Community resilience: toward an integrated approach. Society & Natural Resources, 26(1), 5-20.

Berkes, F. (2012). Understanding Uncertainty and Reducing Vulnerability: Lessons from Resilience Thinking. In C.E. Haque and D. Etkin (eds.) Disaster risk and vulnerability: mitigation through mobilizing communities and partnerships. Canada: McGill-Queen's University Press.

Bruneau, M., Chang, S. E., Eguchi, R. T., Lee, G. C., O'Rourke, T. D., Reinhorn, A. M., ... & Von Winterfeldt, D. (2003). A framework to quantitatively assess and enhance the seismic resilience of communities. Earthquake spectra, 19(4), 733-752.

Burton, H. V., Deierlein, G., Lallemant, D., & Singh, Y. (2017). Measuring the Impact of Enhanced Building Performance on the Seismic Resilience of a Residential Community. *Earthquake Spectra*, *33*(4), 1347-1367.

Çağnan, Z., Davidson, R. A., & Guikema, S. D. (2006). Post-earthquake restoration planning for Los Angeles electric power. Earthquake Spectra, 22(3), 589-608.

Chandra, A., Acosta, J., Howard, S., Uscher-Pines, L., Williams, M., Yeung, D., ... & Meredith, L. S. (2011). Building community resilience to disasters: A way forward to enhance national health security. *Rand health quarterly*, *1*(1).

Chang, S. E., McDaniels, T., Fox, J., Dhariwal, R., & Longstaff, H. (2014). Toward disaster-resilient cities: Characterizing resilience of infrastructure systems with expert judgments. Risk Analysis, 34(3), 416-434.

Chang, S. E., & Rose, A. Z. (2012). Towards a theory of economic recovery from disasters. International Journal of Mass Emergencies and Disasters.

Cimellaro, G. P., Renschler, C. S., Frazier, A., Arendt, L. A., Reinhorn, A. M., & Bruneau, M. (2011). The state of art of community resilience of physical infrastructures. In *Structures Congress 2011* (pp. 2021-2032).

Comes, T., & Van de Walle, B. (2014). Measuring disaster resilience: The impact of hurricane sandy on critical infrastructure systems. ISCRAM, 11, 195-204.

Cutter, S. L., Burton, C. G., & Emrich, C. T. (2010). Disaster resilience indicators for benchmarking baseline conditions. Journal of Homeland Security and Emergency Management, 7(1).

Cutter, S. L. (2016). The landscape of disaster resilience indicators in the USA. *Natural hazards*, *80*(2), 741-758.

Davis, C. A., Mostafavi, A., & Wang, H. (2018). Establishing Characteristics to Operationalize Resilience for Lifeline Systems. *Natural Hazards Review, 19(4)*, 04018014.

Dilmaghani, R. B., Manoj, B. S., & Rao, R. R. (2006, May). Emergency communication challenges and privacy. In Proceedings of the 3rd International ISCRAM Conference.

Eid Mohamed S., and El-adaway Islam H. (2017). Integrating the Social Vulnerability of Host Communities and the Objective Functions of Associated Stakeholders during Disaster Recovery Processes Using Agent-Based Modeling. *Journal of Computing in Civil Engineering, 31(5)*, 04017030.

Ganji, A., & Miles, S. (2018, October). Toward Human-Centered Simulation Modeling for Critical Infrastructure Disaster Recovery Planning. In *2018 IEEE Global Humanitarian Technology Conference (GHTC)* (pp. 1-8). IEEE.

Ganji, A., Orand, M., & McDonald, D. W. (2018). Ease on Down the Code: Complex Collaborative Qualitative Coding Simplified with 'Code Wizard'. *Proceedings of the ACM on Human-Computer Interaction*, *2* (CSCW), 132.

Gonzalez, J. J., Granmo, O. C., Munkvold, B. E., Li, F. Y., & Dugdale, J. (2012, April). Multidisciplinary challenges in an integrated emergency management approach. *In Proceedings of the 9th International ISCRAM Conference.*

Kulig, J. C., Edge, D. S., Townshend, I., Lightfoot, N., & Reimer, W. (2013). Community resiliency: Emerging theoretical insights. *Journal of Community Psychology*, *41*(6), 758-775.

Kumar, S., Diaz, R., Behr, J. G., and Toba, A.-L. (2015). Modeling the effects of labor on housing reconstruction: A system perspective. *International Journal of Disaster Risk Reduction*, 12, 154–162.

Labaka, L., Hernantes, J., Comes, T., & Sarriegi, J. M. (2014, May). Defining policies to improve critical infrastructure resilience. In *ISCRAM*.

Laugé, A., Hernantes, J., and Sarriegi, J. M. (2015). Critical infrastructure dependencies: A holistic, dynamic and quantitative approach. *International Journal of Critical Infrastructure Protection*, 8, 16–23.

Longman, M., & Miles, S. B. (2019). Using discrete event simulation to build a housing recovery simulation model for the 2015 Nepal earthquake. *International Journal of Disaster Risk Reduction, 101075*.

Lee, L., Mitchell, J., Wallace, W., (2007, November). Restoration of Services in Interdependent Infrastructure Systems: A Network Flows Approach. *IEEE Transactions on Systems, Man, and Cybernetics, Part C (Applications and Reviews)*, *Volume: 37 , Issue: 6*.

Maitland, C., Ngamassi, L., & Tapia, A. (2009, May). Information management and technology issues addressed by humanitarian relief coordination bodies. *In Proceedings of the 6th International ISCRAM Conference.*

Miles, S. B., Burton, H. V., & Kang, H. (2018). Community of practice for modeling disaster recovery. *Natural Hazards Review, 20(1)*, 04018023.

Miles, S. B. (2015). Foundations of community disaster resilience: Well-being, identity, services, and capitals. *Environmental Hazards*, *14*(2), 103-121.

Miles, S. B., and Chang, S. E. (2011). ResilUS: A Community Based Disaster Resilience Model. *Cartography and Geographic Information Science*, 38(1), 36–51.

Nejat Ali, and Ghosh Souparno. (2016). LASSO Model of Postdisaster Housing Recovery: Case Study of Hurricane Sandy. *Natural Hazards Review*, 17(3), 04016007.

NIST (National Institute of Standards and Technology). (2016). Community Resilience Planning Guide, Washington, DC. http://www.nist.gov/el/resilience/guide.cfm

Norris, F. H., Stevens, S. P., Pfefferbaum, B., Wyche, K. F., & Pfefferbaum, R. L. (2008). Community resilience as a metaphor, theory, set of capacities, and strategy for disaster readiness. *American journal of community psychology*, *41*(1-2), 127-150.

OSSPAC (Oregon Seismic Safety Policy Advisory Commission). (2013). The Oregon Resilience Plan. Salem, OR: Oregon Seismic Safety Policy Advisory Committee. February 2013. Accessed on June 19, 2018.

Ouyang, M., and Dueñas-Osorio, L. (2012). Time-dependent resilience assessment and improvement of urban infrastructure systems. *Chaos: An Interdisciplinary Journal of Nonlinear Science*, 22(3), 033122.

Poland, C. (2009). The resilient city: Defining what San Francisco needs from its seismic mitigation polices. *San Francisco Planning and Urban Research Association report, San Francisco, CA, USA*.

Ramachandran, V., Long, S. K., Shoberg, T., Corns, S., & Carlo, H. J. (2015). Framework for modeling urban restoration resilience time in the aftermath of an extreme event. Natural Hazards Review, 16(4), 04015005.

Rubim, I., Borges, R.S., (2017, May). The Resilience and Its Dimensions. In *ISCRAM*.

Semaan, B., & Hemsley, J. (2015, May). Maintaining and Creating Social Infrastructures: Towards a Theory of Resilience. In *ISCRAM*.

Scholl, H. J., and Carnes, S. L. (2017). Managerial Challenges in Early Disaster Response: The Case of the 2014 Oso/SR530 Landslide Disaster. 12.

Tabucchi, T., Davidson, R., & Brink, S. (2010). Simulation of post-earthquake water supply system restoration. *Civil Engineering and Environmental Systems*, *27*(4), 263-279.

Turoff, M., Bañuls, V. A., Plotnick, L., Hiltz, S. R., and Ramírez de la Huerga, M. (2016). A collaborative dynamic scenario model for the interaction of critical infrastructures. *Futures*, 84, 23–42.

van Laere, J., Berggren, P., Gustavsson, P., Ibrahim, O., Johansson, B., Larsson, A., Lindqwister, T., Olsson, L., and Wiberg, C. (2017). Challenges for critical infrastructure resilience: cascading effects of payment system disruptions. ISCRAM, 281–292.

WASSC (Washington State, Seismic Safety Committee). (2012). Resilient Washington State: A Framework for Minimizing Loss and Improving Statewide Recovery after an Earthquake. Olympia, WA: State of Washington Emergency Management Council Seismic Safety Committee. November 2012.